\documentclass[]{aa} 

\newcommand{\ltsim}{\protect\raisebox{-0.5ex}{$\:\stackrel{\textstyle <}
{\sim}\:$}}
\newcommand{\mlp}{\ensuremath{\alpha_{\mathrm{MLT}}}}
\newcommand{\kms}{\,$\mathrm{km\, s^{-1}}$}

\usepackage{txfonts}
\usepackage{graphicx}
\usepackage[authoryear]{natbib}

\bibpunct{(}{)}{;}{a}{}{,} 

\begin{document}

\title{Solar twins in M67
\thanks{Based on observations collected at the ESO VLT, Paranal Observatory, Chile, program 278.D-5027(A).}
} 
  

\author{L. Pasquini\inst{1} \and K. Biazzo\inst{1,2} \and P. Bonifacio\inst{3,4,5} \and 
S. Randich\inst{6} \and L. Bedin \inst{7}}

\offprints{L. Pasquini, \email{lpasquin@eso.org}}

\institute{ESO - European Southern Observatory, Karl-Schwarzschild-Strasse 3, 85748 Garching bei M\"unchen, Germany
  \and Istituto Nazionale di Astrofisica, Osservatorio Astrofisico di Catania, Via S. Sofia 78, 95123 Catania, Italy
  \and CIFIST, Marie Curie Excellence Team
  \and GEPI, Observatoire de Paris, CNRS, Universit\'e Paris Diderot, Place Jules Janssen 92190 Meudon, France
  \and Istituto Nazionale di Astrofisica, Osservatorio Astronomico di Trieste, Via Tiepolo 11, 34143 Trieste, Italy
  \and Istituto Nazionale di Astrofisica, Osservatorio Astrofisico di Arcetri, Largo E. Fermi 5, 50125 Firenze, Italy
  \and Space Telescope Science Institute, 3700 San Martin Drive, Baltimore, MD 21218
}

\date{Received  / Accepted }

\abstract
{The discovery of true solar analogues is fundamental for a better understanding of the Sun and of the solar system. 
Despite a number of efforts, this search has brought only to limited results among field stars. The open cluster M67 
offers a unique opportunity to search for solar analogues because its chemical composition and age are very similar 
to those of the Sun.}
{We analyze FLAMES spectra of a large number of M67 main sequence stars to identify solar analogues in this cluster.} 
{We first determine cluster members which are likely not binaries, by combining proper motions and radial velocity 
measurements. We concentrate our analysis on the determination of stellar effective temperature, using analyses 
of line-depth ratios and H$\alpha$ wings, making a direct comparison with the solar spectrum obtained with the same 
instrument. We also compute the lithium abundance for all the stars.}
{Ten stars have both the temperature derived by line-depth ratios and H$\alpha$ wings within 100 K from the 
Sun. From these stars we derive, assuming a cluster reddening $E(B-V)=0.041$, the solar colour 
$(B-V)_\odot=0.649\pm0.016$ and a cluster distance modulus of 9.63. Five stars are most similar (within 60 K) to 
the Sun and candidates to be true solar twins. These stars have also a low Li content, comparable to the 
photospheric abundance of the Sun, likely indicating a similar mixing evolution.}
{We find several candidates for the best solar analogues ever. These stars are amenable to further spectroscopic 
investigations and planet search. The solar colours are determined with rather high accuracy with an independent 
method, as well as the cluster distance modulus.}

\keywords{ stars: fundamental parameters -- open clusters and associations: individual: M67 -- stars: late-type}
   
\titlerunning{True solar analogues in M67}
\authorrunning{L. Pasquini et al.}
\maketitle

\section{Introduction}
\label{sec:Intro}

The specificity of the Sun and of our solar system have been the subject of active investigation over the last 5 
decades. How typical is the Sun for a star of its age, mass, and chemical composition? How typical is that 
solar-type stars host planetary systems? Are they similar at all to ours? 

The quest to find stellar analogues to the Sun has been going on for a long time (for an extensive review see, 
e.g., \citealt{Cayrel1996}), and it stems from the poor knowledge we have of the Sun when seen `as a star' and 
from how typical the Sun is for a G2 type star, for its age, chemical composition, population. It is, however, 
after the discovery of the first exo-planets (\citealt{MayQue1995}) that this quest became even more compelling, 
because to find stars similar to our own would allow us to answer to fundamental questions related to the origin 
of the solar system, the frequency of planetary systems similar to ours, and eventually the formation of life in 
other exo-planetary systems (\citealt{Cayrel1996}). The need to identify in the night sky solar proxies to be used 
for spectroscopic comparison is also diffuse, in particular for the analysis of small solar system bodies 
(B\"ohnhardt, private communication). 

Among the most recent results in this research, \cite{Melendez2006} used high resolution, high signal-to-noise 
ratio Keck spectra to show that HD 98618 is a very close solar twin, and \cite{King2005} proposed HD 143436 after 
analyzing 4 stars pre-selected from literature. These stars seem to compare well with the best known solar twin, HR 6060, 
first analyzed by \cite{PorSil1997}, and subsequently confirmed by \cite{SoubTri2004}, who made a comparative 
study of several hundreds of ELODIE spectra. Finally, \cite{Melendez2007} have shown HIP~56948 to be the best 
solar twin known to date both in stellar parameters and in chemical composition, including a low lithium abundance.

The open cluster M67 is a perfect target to search for solar analogues. Recent chemical analyses 
(\citealt{Tau2000, Randich2006, Pace2008}), show that this cluster has a chemical composition (not only Fe, but 
also all the other elements) extremely similar to the solar one, as close as allowed by the high precision of the 
measurements. The analysis resulted in [Fe/H]=$-$0.03$\pm$0.03 for \cite{Tau2000}, [Fe/H]=0.03$\pm$0.01 for 
\cite{Randich2006}, and [Fe/H]=0.03$\pm$0.03 for \cite{Pace2008}. 

There are other two additional characteristics which make M67 strategical. The first one is that all the 
determinations of age give for this cluster an age encompassing that of the Sun (3.5-4.8 Gyr; 
\citealt{Yadav2008}), while the age determination for field stars is always uncertain. The second characteristic 
is that M67 is among the very few clusters showing Li depleted G stars (\citealt{Pasquini1997}). This is an 
important point because, as pointed out by \cite{Cayrel1996}, even if many stars appear to have most characteristics 
similar to the Sun, their Li abundance is usually 10 times higher than in our star. Since Li is likely an indicator 
of the complex interaction taking place in the past between the stellar external layers and the hotter interior, 
the choice of stars which also share the same Li abundance with the Sun is an additional property to pinpoint the 
true analogues. 

In our opinion, the search of analogues to the Sun and to the solar system can be well performed in open clusters 
(OCs), which show a homogeneous age and chemical composition, common birth and early dynamical environment. As 
a consequence, they provide an excellent laboratory for investigating the physics of solar stars and of planetary 
system evolution, besides being excellent probes of the structure and evolution of the Galactic disk. 

\begin{figure}
\centering
\includegraphics[width=9cm]{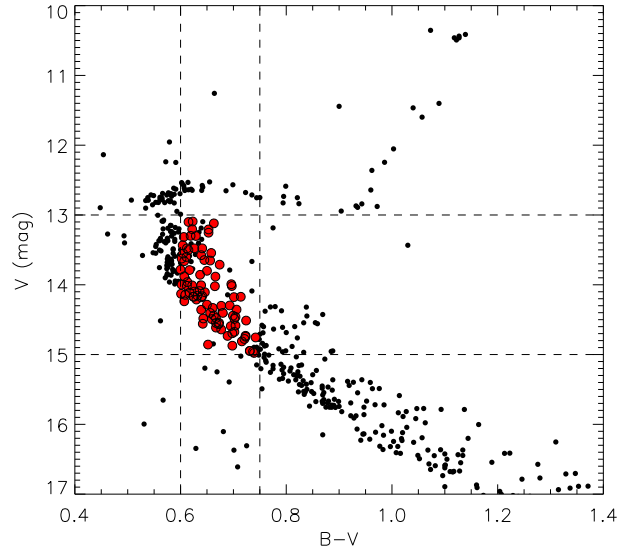}
\caption{Portion of the colour-magnitude diagram of M67.
Our selected targets encompass the solar colour, and are high probability proper motion (\citealt{Yadav2008}) 
and radial velocity single members. The red points refer to the stars observed with FLAMES/GIRAFFE in three nights.} 
\label{fig:cmd_M67}
\end{figure}

M67 is a rich cluster, therefore it provides us with the opportunity to find many stars candidates sharing similar 
characteristics, and not only one. This is fundamental to obtain some meaningful statistics, and the cluster hosts 
many main sequence (MS) stars of mass around the solar mass, which form a continuous distribution 
(Fig.~\ref{fig:cmd_M67}). 

Finding several solar analogues in M67 will also help in providing an independent estimate of the solar colors, 
a quantity which still suffers of some relevant uncertainty (see, e.g., \citealt{Holmberg2006}), as well as an 
independent estimate of the distance modulus of the cluster.

The present paper is the culmination of a work, which involved the chemical determination of this cluster 
(\citealt{Randich2006,Pace2008}), photometry and astrometry (\citealt{Yadav2008}) to obtain membership, and 
FLAMES/GIRAFFE high resolution spectroscopy to clean this sample from binaries, and to look for the best solar 
analogues using the line-depth ratios method (\citealt{GrayJoha1991,Biazzo2007}) and the wings of the H$\alpha$ 
line (\citealt{CayBen1989}) to determine accurate temperatures with respect to the Sun. In addition, the Li line 
is used to separate Li-rich from Li-poor stars. 

\section{Observations and data reduction}
\label{sec:Obs}
We obtained 2.5 hours in three observing nights in service mode with the DDT program 278.D-5027(A). Observations 
were carried out with the multi-object FLAMES/GIRAFFE spectrograph at the UT2/Kueyen ESO-VLT (\citealt{Pasquini2002}) 
in MEDUSA mode\footnote{This is the observing mode in FLAMES in which 132 fibers with a projected diameter on the 
sky of 1\farcs{2} feed the GIRAFFE spectrograph. Some fibers are set on the target stars and others on the sky 
background.}; we were able to observe 90 targets (Table~\ref{tab:total_targets}). The employed setting was HR15N 
with central wavelength 665.0 nm, which with a wavelength range between 644 and 682 nm covers simultaneously 
the H$\alpha$ and the \ion{Li}{i} resonance doublet at 670.8 nm with a resolution of R$\sim$17\,000. Three separate 
exposures were obtained to be able to identify short and intermediate period binaries by comparing the radial 
velocities at different epochs; the combined spectra have a typical signal-to-noise ($S/N$) ratio of 
80-110/pixel. 

We selected from the catalog of \cite{Yadav2008} the main sequence stars ($13\fm0\ltsim V\ltsim15\fm0$) with 
$B-V$ close to that of the Sun ($\approx$0.60--0.75) with the best combination of proper motions parameters, that 
is a membership probability superior to 60\%, and exclusion of candidates with a proper motion larger than 6 mas/yr 
with respect to the average cluster members. Full details about proper motion errors and selection criteria 
can be found in the original \cite{Yadav2008} work.

The log of the observations is given in Table~\ref{tab:observations}. The observations were reduced using the 
ESO-GIRAFFE pipeline.

Radial velocities were measured using the IRAF\footnote{IRAF is distributed by the National Optical Astronomy 
Observatory, which is operated by the Association of the Universities for Research in Astronomy, inc. (AURA) under 
cooperative agreement with the National Science Foundation.} package FXCOR, which cross-correlates the observed 
spectrum with a template. As a template we used a solar spectrum acquired with FLAMES/GIRAFFE. Finally, the 
heliocentric correction was applied. The typical error for our single measurement is around 0.4\kms. The three 
spectra/star were finally co-added to perform the spectroscopic determination of temperature and 
lithium abundance (see Sections \ref{sec:teff} and \ref{sec:lithium}). 

We note that the GIRAFFE solar spectrum\footnote{http://www.eso.org/observing/dfo/quality/GIRAFFE/pipeline/solar.html}, 
taken with the same setup of our observations, is used through this work for spectroscopic comparison with the stars 
and the synthetic spectra. The solar spectrum has been obtained by averaging most of the GIRAFFE 
spectra (some show clear flat field problems and have not been used) and it has a nominal $S/N$ ratio above 400.

\begin{table}  
\caption{Log of the observations.}
\label{tab:observations}
\begin{center}  
\begin{tabular}{cccccc}
\hline
\hline
$\alpha$      & $\delta$  &  Date      &  UT      &  $t_{\rm exp}$  & DIMM seeing \\
              &           & (d/m/y)    & (h:m:s)  &  (s)            & (arc sec)    \\ 
\hline
132.875     & 11.833   & 06/02/2007 & 06:24:29 & 2200		 & 0.6         \\
132.875     & 11.833   & 11/02/2007 & 04:01:08 & 2200		 & 1.1         \\
132.875     & 11.833   & 23/02/2007 & 01:34:55 & 2100		 & 0.9         \\
\hline
\end{tabular}
\end{center}
\end{table}  

\section{Data analysis and membership}
\subsection{Radial velocity}
\begin{figure}
\centering
\includegraphics[width=9cm]{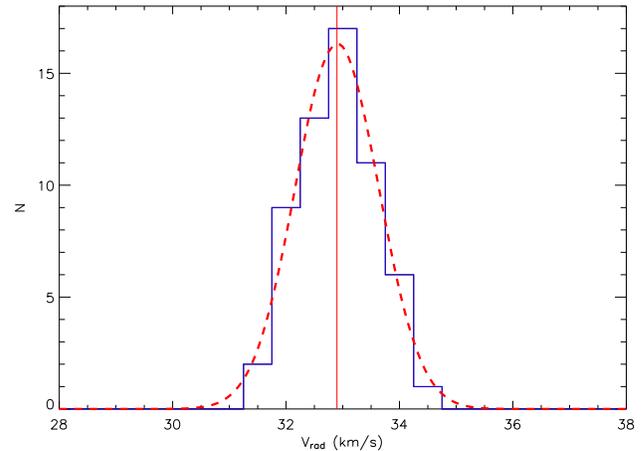}
\caption{Histogram of the radial velocity distribution of the 59 single members selected in M67 (continuous line). 
A Gaussian fit to the member stars distribution is also displayed (dashed line).}
\label{fig:vrad_distribution_sel}
\end{figure}

Out of the 90 stars observed, all selected on the proper motion and membership criteria given above, we found 
that 59 of them are probable single radial velocity (RV) members. We have retained all the stars which show RV 
variations smaller than $\approx$1 km s$^{-1}$ in the three exposures acquired and which have a mean velocity 
within 2 sigma ($\approx$1.8 km s$^{-1}$) from the median cluster RV. In Fig.~\ref{fig:vrad_distribution_sel} the 
histogram of the radial velocity distribution of these stars is shown, together with a Gaussian fit with 
$<V_{\rm rad}>$=32.90 km s$^{-1}$ and a $\sigma=0.73$ km s$^{-1}$. In Table~\ref{tab:targets_selected} the RV values 
are listed for the stars of the final sample, while in Table~\ref{tab:nonmembers} the values of the single RV 
measurements are given for the stars we discarded.

\begin{figure}
\centering
\includegraphics[width=9cm]{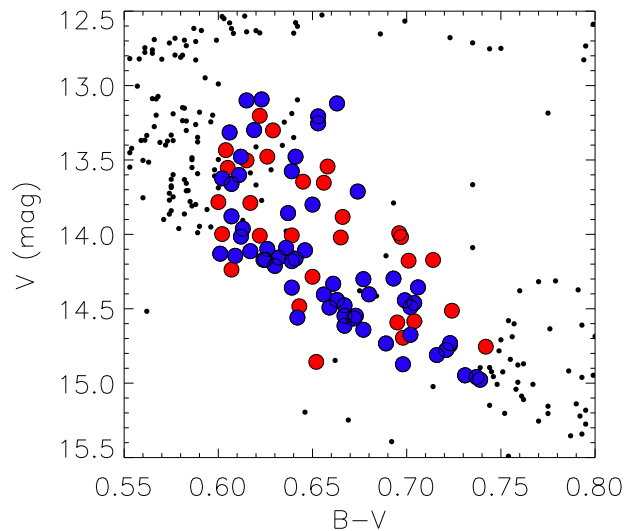}
\caption{Zoom of the M67 CMD, centered on the targets observed. In blue the 59 retained single 
member candidates are shown, in red the stars discarded. As expected by binary contamination, most of the discarded 
stars lay on the bright side of the MS.}
\label{fig:cmdenlarged}
\end{figure}

In Fig.~\ref{fig:cmdenlarged} we show the enlarged portion of the colour-magnitude diagram CMD containing 
the original sample; in this Figure the discarded and the retained stars are indicated with different colours. 
Many of the discarded stars tend to occupy the brighter side of the main sequence, where binaries are indeed 
expected to be present. On the other hand, our procedure still leaves several stars which are apparently above 
the photometric main sequence. This is because the radial velocity measurements are not of superb quality and 
because the time span by the observations is of only 18 days. Long period binaries will not be discovered by our 
three radial velocity observations. We shall see as seven stars clearly stand up also in the Magnitude -- Temperature 
diagram (see Fig.~\ref{fig:Teff_V}) and they are best candidates for binaries of similar mass. We have kept them 
in the sample, and we anticipate that their presence does not influence our analysis or conclusions. 

\subsection{Effective temperature}
\label{sec:teff}
Given that our targets are on the main sequence of a cluster of solar metallicity and age, the critical astrophysical 
parameters for the selection of the best solar analogues is the effective temperature. We have used two spectroscopic 
methods to compute the stellar effective temperature: the line-depth ratios and the H$\alpha$ wings. To calibrate 
these methods we have used a grid of synthetic spectra, computed with SYNTHE from a grid of 1D LTE model atmospheres 
computed with version 9 of the ATLAS code (\citealt{Kurucz1993a,Kurucz1993b}) in its Linux version 
(\citealt{Sbord2004,Sbord2005}). All the models have been computed with the ``NEW'' Opacity Distribution 
Functions (\citealt{CK03}) which are based on solar abundances from \cite{Grevesse1998} with 1\kms\ micro-turbulence, 
a mixing-length parameter~\mlp\ of 1.25 and no overshooting. The grid of synthetic spectra covers the temperature 
range 5450--6300 K with [Fe/H]=0, $\log g$=4.4377, $\xi$=1 km s$^{-1}$ and was degraded to the resolution of the 
FLAMES/GIRAFFE spectra. We stress that for both methods these models are used to quantify the difference between 
the stellar spectra and the solar spectrum. Zero point shifts are most likely present, due, for instance, 
to limitations in the atmospheric models or to not perfect treatment of the H$\alpha$ lines. While
these inaccuracies will reflect in a wrong temperature for the Sun, the difference between the stars and 
the Sun will be much less affected.

\subsubsection{LDR method}
\begin{figure}
\centering
\includegraphics[width=9cm]{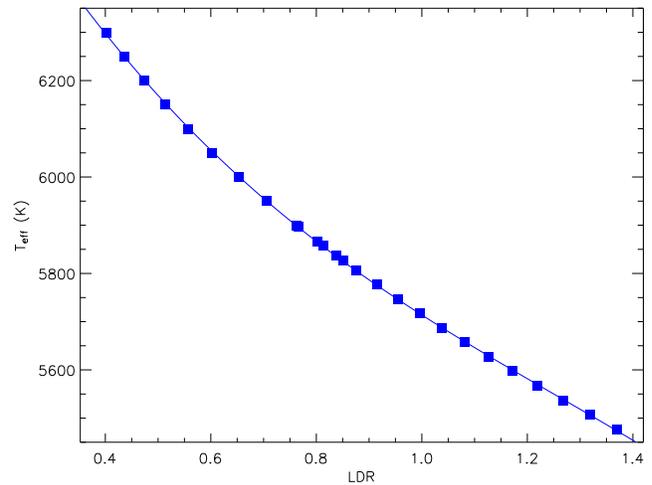}
\caption{Example of LDR$-T_{\rm eff}$ calibration obtained with synthetic spectra. The line-depth ratio is between 
$\lambda$6498.937~\AA~and $\lambda$6516.05~\AA.} 
\label{fig:d6469_d6456}
\end{figure}

It has been demonstrated that for stars with $B-V=0.4-1.5$ line-depth ratios (LDRs) are a powerful temperature 
indicator, capable to resolve temperature differences lower than 10 K 
(\citealt{GrayJoha1991,Catalano2002,Biazzo2007}). Since our stars are within this $B-V$ range, we have 
applied the LDR method to the members previously selected by radial velocity measurements (see 
Table~\ref{tab:targets_selected}). To convert the line-depth ratios of our stars into effective temperature we 
need to calibrate a temperature scale for the measured LDRs. To this purpose we have considered an initial 
sample of about 100 lines of iron group elements (which are usually temperature sensitive) present in the 
spectral range covered by our observations, from which we selected lines with the following characteristics: 
weak (to avoid saturation effects), sensitive to temperature variations, and at the same time well measurable 
in our spectra. The final selection contains six line pairs suitable to apply the LDR method; we measured 
them in the synthetic spectra and then derived an LDR$-T_{\rm eff}$ calibration for each pair. 
In Table~\ref{tab:line_list} we list for these six line pairs the wavelength, the element and the excitation 
potential, as taken from the NIST\footnote{National Institute of Standards and Technology.} Atomic Spectra 
Database Lines, while in Fig.~\ref{fig:d6469_d6456} we show an example of one LDR$-T_{\rm eff}$ calibration. 
The methods for the measurement of the line depth and the related uncertainties are described in 
\cite{Catalano2002} and \cite{Biazzo2007}. In a summary, the lowest seven points in the core of each measured 
line were fitted with a cubic spline and the minimum of this cubic polynomial was taken as the line depth. 
Given the limited $S/N$ ratio and resolution, the errors in each line depth are dominated by the uncertainty of the signal in the continuum. 
We have then measured the LDRs and  derived for each line pair the temperature; 
in Table~\ref{tab:targets_selected} the averaged values for all line pairs are given, where the 
associated uncertainty reflects the scatter obtained by the six measurements.

\begin{table}	
\caption{List of the six line pairs used to derive the stellar temperature.}
\label{tab:line_list}
\begin{center}
\begin{tabular}{lrc}
\hline
\hline
{$\lambda_1/\lambda_2$ }&{  LDR }&{ $\chi_1/\chi_2$ }\\
  (\AA)   &           &  (eV)\\\hline
6469.210/6456.380  & Fe{\sc{i}}/Fe{\sc{ii}}  & 4.84/3.90\\
6498.937/6516.050  & Fe{\sc{i}}/Fe{\sc{ii}}  & 0.96/2.89\\
6608.024/6597.557  & Fe{\sc{i}}/Fe{\sc{i}}   & 2.28/4.80\\
6608.024/6627.540  & Fe{\sc{i}}/Fe{\sc{i}}   & 2.28/4.55\\
6646.932/6627.540  & Fe{\sc{i}}/Fe{\sc{i}}   & 2.61/4.55\\
6646.932/6653.850  & Fe{\sc{i}}/Fe{\sc{i}}   & 2.61/4.15\\
\hline
\end{tabular}
\end{center}
\end{table}

With this method the effective temperature of 
the observed GIRAFFE solar spectrum results in 5792$\pm$27 K, i.e. 15 K higher than the synthetic one (5777 K 
is the theoretical effective temperature of the solar atmosphere; \citealt{Wilson1991}).

We computed the temperature difference $\Delta T^{\rm LDR}$ between the FLAMES/GIRAFFE targets and the 
Sun (as obtained from the six line-depth ratios and the summed spectra of the targets) as a function of the de-reddened 
$B-V$ colour ($E(B-V)=0.041$; \citealt{Taylor2007}). The $\Delta T^{\rm LDR}-(B-V)_0$ relationship for our targets is well 
described by a linear fit, which gives $\Delta T^{\rm LDR}=(-3662.65\pm351.22)\times (B-V)_0+(2410.59\pm216.52)$ 
and an rms of 100 K.

\begin{figure}
\centering
\includegraphics[width=9cm]{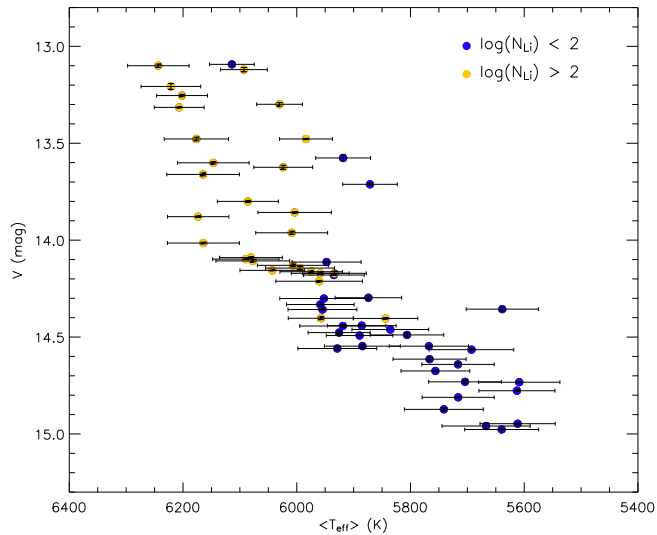}
\caption{$V$ vs. $T_{\rm eff}$ of the probable candidates. The stars have different colours according to 
the LTE lithium abundance. Seven stars depart from the main sequence; they are most likely long period 
binaries which escaped the detection in our observations.
} 
\label{fig:Teff_V}
\end{figure}

In Fig.~\ref{fig:Teff_V} the temperature-magnitude diagram is shown. The two colours of the symbols are referred 
to stars with lower and higher presence of lithium. Seven stars clearly stand out of the main sequence, suggesting 
a parallel binary sequence. They most likely are long period binaries with components of similar mass not detected 
as RV variable by our observations, because of the limited time base of our observations.  



\subsubsection{H$\alpha$ line}
The wings of the H$\alpha$ line profile are very sensitive to temperature, like all the Balmer lines, and depend 
only slightly on metallicity and gravity (\citealt{Cayrel1985,Fuhrmann1993,Barklem2002}). In particular, the 
spectral region in the range between 3 and 5 \AA~from the H$\alpha$ line center is a good effective temperature 
diagnostic (\citealt{CayBen1989}). With respect to the higher members of the series, the H$\alpha$ line has 
considerable less blending in the wings, making easier the placement of the continuum. A further advantage, over 
other members of the series, is that it is rather insensitive to convection and in particular to the adopted 
mixing length parameter \mlp\ in 1D model atmospheres (\citealt{Fuhrmann1993}). Thus, we selected this region 
as temperature indicator and for each star we have compared the H$\alpha$ line profile outside the core 
in our real spectrum against the synthetic profile. The hydrogen line profiles 
are computed in SYNTHE by routine {\tt HPROF4}, which, for resonance broadening uses essentially the 
\cite{AG65,AG66} theory, and for Stark broadening it calls routine {\tt SOFBET}, written by Deane Peterson, 
which uses essentially the theory of \cite{Griem}, with modified parameters, so as to provide a good approximation 
to the \cite{VCS} profiles (F. Castelli, private communication). For further details on the computations of 
hydrogen lines in SYNTHE, see \cite{CK01} and \cite{CC2002}. The dominant broadening for H$\alpha$ is resonance 
broadening, while Stark broadening becomes relevant for higher members of the series. In order to minimize the 
subjectiveness of the measurement, we have quantified the comparison between the synthetic profile and the observed 
one minimizing the rms of the subtraction. Continuum normalization is not easy for such a broad line, however, 
the fact of using a fiber instrument with a large coverage, minimize the subjectiveness of the process and makes 
it quite reproducible. Given the limited $S/N$ ratio of the observations, it is however very difficult to provide a 
realistic estimate of the involved uncertainties. The systematic errors in the effective temperature obtained 
from the H$\alpha$ wings is given by \cite{Gratton2001}, and the errors associated to the method have been 
discussed by, e.g., \cite{B07}, where the dominant source of error for \'echelle spectra has been identified in 
residuals in the correction of the blaze function. The GIRAFFE spectra are fiber-fed and the flat-field is 
obtained through the same optical path as the stellar spectra, thus flat-fielding allows a better removal of the 
blaze function than it is possible for slit spectra. We have estimated for our stars an average error of 
$\pm$100 K. With this method the effective temperature of the Sun results in 5717$\pm$100 K, i.e. 60 K lower 
than the solar real value. We note that the absolute temperature determined in this way depends critically on 
a number of assumptions in the model, and on the adopted broadening theory for H$\alpha$, and these produce a 
zero point shift of the Sun. The relative measurements, which are made with respect to the observed solar 
spectrum, are instead rather insensitive to all the assumptions used to build the synthetic profile.

A linear fit well describes the relationship between the temperature difference $\Delta T^{\rm H\alpha}$ 
of the FLAMES/GIRAFFE targets and the Sun as obtained from the H$\alpha$ wings, and the de-reddened $B-V$ 
colour: $\Delta T^{\rm H\alpha}=(-3811.8\pm283.39)\times (B-V)_0+(2488\pm174.71)$ with an rms of 81 K.
 
The calibrations $\Delta T_{\rm eff}$ vs. $(B-V)_0$ obtained with the two methods agree quite well, 
as shown in Fig.~\ref{fig:deltaT}; they have slightly different slopes which produce a maximum 
difference at the red edge ($(B-V)_0=0.7$) of the sample of 40 K (H$\alpha$ temperatures are cooler). 
These relationships can be used to calibrate stars with metallicity close to solar. Our LDR $T_{\rm eff}$ 
vs. $(B-V)_0$ relationship has almost exactly the same slope of the \cite{Alonso1996} relationship, but it 
is hotter than this by $\sim$60 degrees. As a reference, the \cite{Alonso1996}'s scale produce an 
effective temperature for the Sun of 5730 K for a $B-V=0.63$.
  


\subsection{Lithium}
\label{sec:lithium}
Lithium is an important element because it is easily destroyed in stellar interiors, and its abundance 
indicates the amount of internal mixing in the stars. Lithium in Pop I old solar stars varies by a factor 10 
(\citealt{pas94}) and M67 is one of the few clusters which clearly shows this spread among otherwise similar stars 
(\citealt{Pasquini1997}).

Equivalent widths (EWs) of the lithium line at $\lambda=670.7876$\,nm~were computed using the IRAF task SPLOT; 
from measured EWs we derived Li abundances using the curves of growth (COGs) of \cite{Soderblom1993}. At the 
GIRAFFE resolution the \ion{Li}{i} lines are blended with the \ion{Fe}{i} $\lambda$ 670.744\,nm~line, whose 
contribution to the lithium blend was subtracted using the empirical correction of the same authors. Lithium 
abundances were then corrected for the NLTE effects using the prescriptions of \cite{Carlsson1994}. 

\begin{figure}
\centering
\includegraphics[width=9cm]{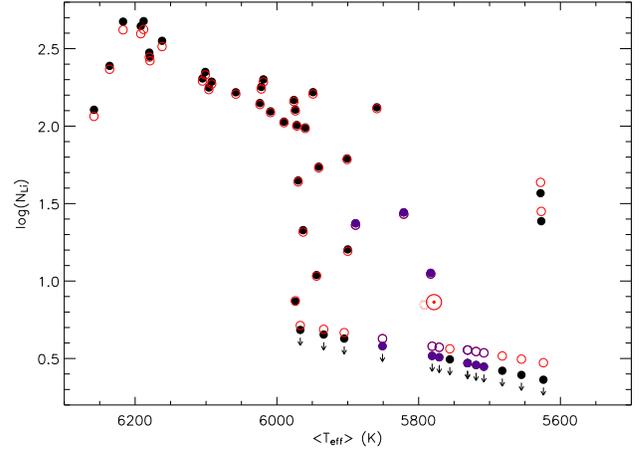}
\caption{Lithium abundance versus $T_{\rm eff}^{\rm LDR}$ for the most probable single stars observed in M67. 
Filled circles: LTE abundances. Open circles: NLTE abundances. The blue points indicate the best solar 
analogues candidates. The position of the Sun is also shown at $\log{N({\rm Li})}^{\rm LTE}=0.842$ and 
$T_{\rm eff}^{\rm LDR}=5792\pm27$ K.} 
\label{fig:EW_Li_BV}
\end{figure}

Fig.~\ref{fig:EW_Li_BV} shows the lithium abundance for the summed spectra of the 59 targets as a function of the 
effective temperature, as derived by the LDR method. The filled symbols refer to the LTE abundance, while the empty 
ones represent the non-LTE abundance. The difference between LTE and non-LTE values is minor. The blue points are 
the solar twins (see Section \ref{sec:solar_analogues}). The position of the Sun is shown at 
$\log{N({\rm Li})}^{\rm LTE}=0.84$ and $T_{\rm eff}^{\rm LDR}=5792\pm27$ K (see Section \ref{sec:teff}). The lithium 
abundance is listed in Table \ref{tab:targets_selected}. Several solar twin candidates have Li abundances 
which are comparable with the Sun, whose value is 0.84 as measured by us on the GIRAFFE spectrum (see above), and 
1.0 as measured in high resolution solar atlas (\citealt{Muller75}). In most investigations the error associated to 
the effective temperature is usually the dominant one (100 K correspond to about 0.1 dex in $\log N{\rm (Li)}$), 
but in this case, since we have good $T_{\rm eff}$ determination, but limited resolution and $S/N$ ratio, the 
uncertainty in the abundance associated to the equivalent width measurements is not negligible. The expected 
uncertainty in the measured lithium equivalent widths has been estimated from \cite{Cayrel88}'s formula:
$$\sigma_{\rm EW_{\rm Li}}=\frac{1.6}{S/N} \sqrt{FWHM \times \delta x}$$
where $S/N$ is the signal-to-noise ratio per pixel, $FWHM$ is the full width of the line at half maximum, 
and $\delta x$ the pixel size. The predicted accuracy, $\sigma_{\rm EW_{\rm Li}}$, is 3.0 m{\AA} for a typical 
$S/N$ ratio of 80 and of 1.6 m{\AA} for a $S/N$ ratio of 150. However, it should be noted that this formula 
neglects the uncertainty on the continuum placement. We estimate that, using homogeneous procedures for the 
determination of the continuum and the line widths, the statistical error for the weak lithium line is of the 
order of 2-3 m{\AA}, depending on the $S/N$ ratio of the co-added spectrum. This will correspond to an asymmetric 
error $\Delta\log N{\rm (Li)}\sim^{+0.3}_{-0.8}$ dex for a star with a line as weak as the Sun, and of 
$\pm 0.04$ dex for a star with a $\log N{\rm (Li)}$=2.2. Since the line is weak, the error (in percentage) is 
inversely proportional to the line strength. Given the errors, all the stars with upper limits in our sample 
may have a Li comparable to the solar one.  

After the early works on M67, several Li surveys have been carried out of additional clusters well sampling 
the age metallicity space (\citealt{Randich2008} and references therein): out of nine clusters older than the 
Hyades with available Li measurements, only two, besides M67, show a significant dispersion. The latter seems 
to be an exception, rather than a rule and its occurrence does not depend on age, nor on metallicity, nor on 
global cluster parameters.

In this context, the novel result of our analysis and, in particular, of the careful selection and cleaning 
of the sample as well as of the precise effective temperature determination, is that the large spread is 
clearly present only for stars cooler than $\sim$ 6000~K. Stars warmer than 6200~K seem to show a decay, 
probably indicating the red side of the ``Li-gap", while stars in the 6000$\leq T_{\rm eff}\leq6200$~K 
do not show any major scatter.

It is now well ascertained on empirical grounds that, in order to explain the MS Li depletion in 
solar-type stars, an extra or non-standard mixing mechanism must be at work. No consensus so far has been 
found on the nature of this mechanism; it is nevertheless clear that, whatever this process is, it must be 
driven by an additional stellar parameter besides mass and chemical composition. The presence of the Li spread 
indeed indicates that this parameter must vary from star to star, and that, depending on it, some stars 
(including the Sun) undergo a much more efficient mixing than others, while the absence of a dispersion for 
stars warmer than 6000~K suggests that this parameter is more uniform among F-type stars. 

Recent modeling have had some success in reproducing the solar Li abundance by using fairly complex models 
which include internal gravity waves (\citealt{chata05}), however, those models are not able to reproduce the 
observed evolution of Li with age, and, in particular, the ``plateau'' in Li abundances at old ages 
(\citealt{Randich2008}). We are not aware of similar models for a grid of masses, to be compared to our 
observations; since the number of possible parameters which influence the Li evolution is very large (depth 
of convective zone, initial rotation, magnetic field, mass losses and torques just to mention a few), we cannot 
really predict at present why the extra mixing takes places at a given $T_{\rm eff}$ in M67 stars. 


\section{Solar analogues}
\label{sec:solar_analogues}
\begin{figure}
\centering
\includegraphics[width=9cm]{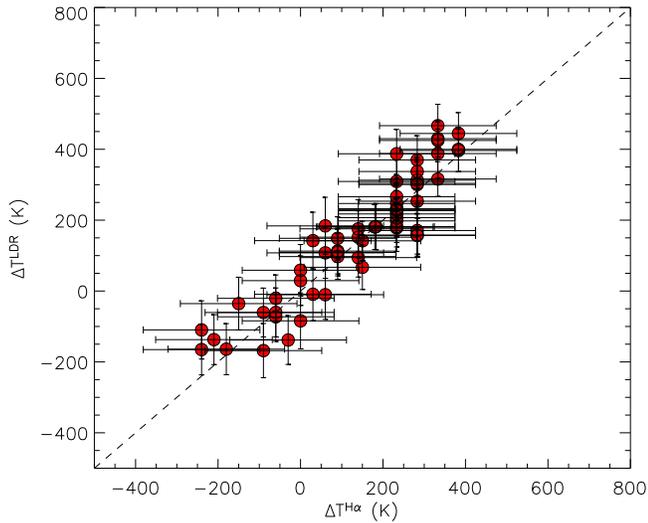}
\caption{$\Delta T^{\rm LDR}$ as a function of $\Delta T^{\rm H\alpha}$ for the 59 probable single member stars.} 
\label{fig:deltaT}
\end{figure}

With the aim to find the best solar analogues, we have compared $\Delta T^{\rm LDR}$ to $\Delta T^{\rm H\alpha}$ 
(Fig.~\ref{fig:deltaT}).

There are in our sample 10 stars for which both the $T_{\rm eff}^{\rm LDR}$ and $T_{\rm eff}^{\rm H\alpha}$ are 
within 100 K of the solar values (285, 637, 1101, 1194, 1303, 1304, 1315, 1392, 1787, 2018). We will use these stars 
to find the best solar analogues and our best evaluation for the solar colours and the cluster distance. These stars 
are indicated in bold face in Table \ref{tab:targets_selected}. 

The average difference between these 10 stars and the solar $T_{\rm eff}^{\rm LDR}$ is of $-$13 K (with a sigma of 
60 K), while the average $T_{\rm eff}^{\rm H\alpha}-T_{\rm eff}^{\rm H\alpha,\odot}$ is =$-$9 K, with a sigma of 
58 K. The average characteristics of these 10 stars should therefore well represent the solar values. 

The average $B-V$ of the ten analogues is $<B-V>=0.692$ ($\sigma=0.020$), their average magnitude 
is $<V>=14.583$ mag ($\sigma=0.147$), and the $<V-I>=0.754$ ($\sigma=0.025$). The $B-V$ spread is larger 
than the formal errors in the photometry, indicating a possible real spread in the stellar characteristics. 
This is not surprising because, formally, these stars may span a range of up to 200 K in temperature. 

If we take our results of Table~\ref{tab:total_targets}, two stars (637 and 1787) have both $T_{\rm eff}$ 
determinations within 50 K from the solar values, and three additional ones (285, 1101, 1194) within 60 K; these 
5 stars are overall the closest to the Sun, with nominal effective temperatures derived with both methods differing 
less than 60 K from the solar one. Their average magnitude (14.557 mag) is very similar to what found for the full 
subsample of 10, as well as their average $B-V$ colour (0.688) just 0.007 magnitude bluer than the whole subsample.
 
All the data in our possession indicate that some of these stars have a metallicity very close (within 0.03 dex; 
note that also their Li abundance is comparable; see previous Section) to the Sun, they have a very similar 
temperature (within 50 K), as well as a comparable age to the Sun, and they are true main sequence stars. To the 
best of our knowledge they are the best candidates in M67 to be the closest analogues to our star. 

\begin{figure}
\centering
\includegraphics[width=9cm]{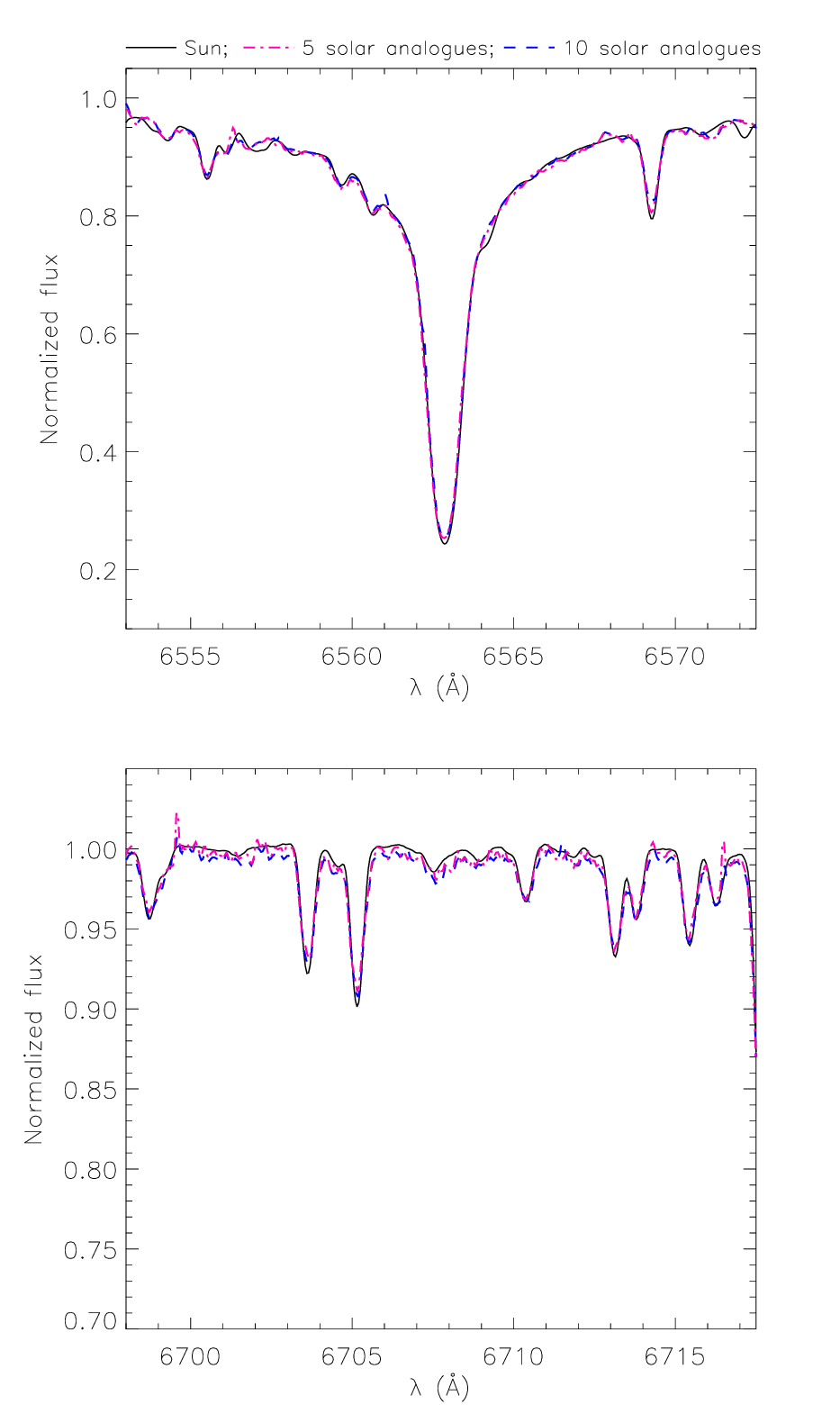}
\caption{Comparison between the GIRAFFE solar spectrum and the spectra of the 10 and 5 best solar analogues. 
The H$\alpha$ and lithium regions are shown in the upper and bottom panels, respectively.} 
\label{fig:solar_twins_comp_Sun}
\end{figure}

In Figure \ref{fig:solar_twins_comp_Sun} we compare the GIRAFFE spectrum of the Sun with the sum of the spectra of 
the 10 best stars analogues and of the 5 best analogues in a portion of the spectra which includes H$\alpha$ and in 
another including the Li lines. The extremely small difference between the solar spectrum and these co-added spectra 
confirm quantitatively the very close resemblance of these stars to the Sun.

At the request of the referee we performed also a direct comparison between the solar spectrum and the 
spectrum of our solar analogs. We used a $\chi^2$ minimization using a Doppler shift and a re-adjustment of the 
continuum of the stars of M67 as free parameters, in order to match the observed spectra to the observed GIRAFFE 
solar spectrum. The reduced $\chi^2$ of the fit, or the associated probability, then provides a mean to rank the 
M67 stars. We restricted the comparison to a range of 10 \AA\ centered on H$\alpha$. While in the fitting of the 
synthetic spectra the core of the line was excluded from the fitting range, it was here included. The LTE synthetic 
spectra fail to reproduce the core of H$\alpha$ due to the presence of a chromosphere (absent in the model 
atmospheres employed by us) and to NLTE effects. Instead, the sought-for solar analogs must behave exactly like 
the Sun, including in the core of H$\alpha$. With this method the three M67 stars which are most similar to the 
Sun are the stars 1194, 1101 and 637. The result is thus very similar to what obtained by comparing the observed 
spectra to synthetic spectra, confirming that the stars we selected are very similar to the Sun. We prefer the 
method based on synthetic spectra, since the direct comparison to the solar spectrum is affected by the noise 
present in the latter.

\section{Solar Colour}
\label{sec:colours}
We would like finally to use the observed $B-V$ colour of the solar analogues to derive in an independent way 
the $B-V$ colour of the Sun, and this requires to evaluate the cluster reddening.

The reddening towards M67 has been evaluated by many authors in the last 50 years, and a thorough discussion is 
given by \cite{Taylor2007}. M67 reddening is evaluated by this author in $E(B-V)=0.041\pm0.004$, which is 
accidentally the same value obtained by \cite{An2007} as the average point of the traditionally accepted 
range for the cluster. We will therefore adopt this value. This implies that the de-reddened colour for the 
average of our 10 solar analogues is $(B-V)_{\odot}=0.651$.
 
The value is in excellent agreement with what found by inverting the fit of all the stars using the $T^{\rm LDR}$, 
which would have predicted $(B-V)_{\odot}=0.659$ and the value obtained by inverting the fit of 
$T^{\rm H{\alpha}}$, which would give $(B-V)_{\odot}=0.651$. 
 
It is not simple to evaluate a realistic error estimate for this colour. This should include: the spread (0.020 mag) 
around our determination, the uncertainty in the cluster reddening, plus other systematics originating from stellar 
evolution and photometry. We evaluate the evolutionary effects by investigating 
the expected variations of the solar color  with age and metallicity by using 
evolutionary models. Photometric uncertainties are estimated by comparing 
Yadav et al. (2008) photometry with what obtained for this cluster by other groups. 
 
We use the tracks from \cite{Girardi2000} for analyzing differential evolutionary effects. Because our stars 
have a similar effective temperature to the Sun, age has no influence: for stars younger than 5 Gyr of solar $T_{\rm eff}$, 
the visual absolute magnitudes and colours do not change in any appreciable way. If M67 were younger than the Sun, 
the only effect would be that the masses of our stars were higher, by about 1\%, than the solar one, but no 
difference is predicted in magnitude or colours. The other source of systematic uncertainty is the possibility 
that metallicity is not exactly solar. In this case, for a fixed effective temperature, we do expect that a star 
more metal rich by 0.05 dex would be slightly brighter (0.08 magnitude in $V$) and slightly redder (0.01 mag) than 
the Sun.  
 
Our photometry is taken by \cite{Yadav2008}, 
which was calibrated on \cite{Sandquist2004}. In order to check for 
photometric systematic errors, we have compared our values for the 10 best analogues with \cite{Montgomery1993}, 
finding that using their photometry an average $(B-V)_0$=0.650 would have been found, i.e. only 
1 mmag bluer than our value. \cite{Sandquist2004} made on the other hand a general comparison between his 
colours and those of \cite{Montgomery1993}, finding an overall zero point shift in $B-V$ of 8 mmag (the 
Sandquist's $B-V$ are bluer than the Montgomery ones). The fact that the agreement for these 10 solar stars is 
better than this systematic shift might be due to a statistical fluctuation, it is on the other hand quite common 
that calibrations agree at best for solar stars. We will nevertheless consider a 0.008 uncertainty in the colour 
as introduced by the adopted photometry.   
   
The simple average of the most recent estimate of the M67 metallicity gives [Fe/H]=0.01. 
This  would  require a correction of  2 mmag towards the blue for the solar colors derived from the 
M67 stars to compensate for their higher metallicity. We conclude that a solar 
$(B-V)_{\odot}=0.649\pm0.016$ is our present best estimate. The uncertainties associated are 0.006 magnitudes 
given by the spread of our solar analogues divided by the square root of number of our solar analogues 
(namely 10); a 0.007 magnitudes given by a generous uncertainty in the cluster metallicity ($\pm$0.03). Zero 
points uncertainties in photometry are the dominant source and they account for 0.008. All these errors are 
summed quadratically. To this, the uncertainty in the cluster reddening determination, which is assumed to be 
0.004 mag (\citealt{Taylor2007,An2007}), is linearly added. 

An additional hidden source of systematic effects, which might add a bias towards redder colours, might be 
present, and this is the presence of unidentified binaries. A typical red, faint companion will make the stars 
to appear slightly brighter and slightly redder than what they should be, still influencing very little the 
spectroscopic $T_{\rm eff}$ determination. Given our radial velocity selection, only a few binaries should be left 
in our sample, and with low mass objects. We cannot quantitatively account for their presence, but 
we shall keep this possibility in mind.  
 
The $B-V$ value found is somewhat in the middle between the majority of the `old' determinations (see Table~2 of 
\citealt{Barry1978}, which found $<(B-V)_{\odot}>$=0.667 averaging most of previous measurements), and the 
most recent determinations, which, as summarized, for instance, by \cite{Holmberg2006}, tend to find 
$(B-V)_{\odot}$ in the range between 0.62 and 0.64. None of these results are formally in disagreement with ours, 
but we can exclude the estimates at the edges of the distribution. 

We think that this estimate is very robust, because our results are based on a very {\it few} steps and assumptions. 
We assume that the metallicity of M67 is essentially solar, and this fact is agreed on by all latest works. We 
determine the $T_{\rm eff}$ in a differential way from the Sun, on spectra taken with the same instrument, and 
using two sensitive methods (line-depth ratios and H$\alpha$ wings). We prove that the stars are indeed very close 
to the Sun showing how their spectra overlap with the solar one. The stars observed are still on the main sequence. 
 
\section{Cluster distance}

The average $V$ magnitude of the 10 solar analogues is 14.583 mag, which must be corrected for reddening: 
$A_{\rm V}$=3.1$\times$0.041=0.127, implying a de-reddened $V$ magnitude of 14.456. With a solar absolute 
$V$ magnitude of 4.81 (\citealt{BCP}), the distance modulus of M67 is of 9.65. As mentioned in the previous 
Section, a correction might be needed if the metallicity differs substantially from the solar one (of up to 
0.08 magnitudes for [Fe/H]=0.05, but we consider such a large difference in metallicity very unlikely). 
A correction of 0.002 mag, corresponding to a metallicity of [Fe/H]=0.01 (used in the previous Section) would 
bring to a distance modulus of 9.63.
 
This determination is in excellent agreement with two recent determinations: \cite{An2007} who estimate a 
distance modulus of 9.61 and \cite{Sandquist2004}, who find 9.60, both using the same reddening we adopted.  

The associated error given by the spread around the average magnitude is of 0.060 (i.e. 0.19/$\sqrt{10}$) 
magnitudes. Other sources of uncertainty in the distance modulus will be given by the error in reddening, 
which accounts for 0.012 magnitudes, and by the uncertainty on [Fe/H]. If we assume an error on [Fe/H] of 0.03 
dex, this accounts for 0.05 magnitudes in the distance modulus. Summarizing, our best estimate of the distance 
modulus is: 9.63$\pm0.06_{\rm stat}\pm0.05_{\rm sys}$.

A full comparison of our estimate with those present in literature is beyond the scope of this work. We find however 
remarkable the agreement between our distance estimate and the ones of \cite{Sandquist2004} and \cite{An2007} in 
particular when considering that our method is independent with respect to theirs. 

\section{Conclusion}

By using selected observations with FLAMES/GIRAFFE at the VLT, we have made a convincing case that the open cluster 
M67 hosts a number of interesting potential solar twins, and we have identified them. We have computed spectroscopic 
accurate effective temperatures for all the stars with two methods. The color-temperature relationships we derive 
can be used to determine temperatures for MS solar-metallicity stars. 

By computing the average solar twin colours, we have obtained a precise estimate of the solar $(B-V)$: 
$(B-V)_{\odot}=0.649\pm0.016$. 

By averaging the magnitude of the solar twins, we have determined an accurate distance modulus for M67: 
9.63$\pm0.06_{\rm stat}\pm0.05_{\rm sys}$, which is in excellent agreement with the most recent estimates, 
which were based on different, independent methods and data sets.  

We have determined for all the stars Li abundances, confirming the presence of a large Li spread among the solar 
stars of this cluster, but showing for the first time, that the Li extra-depletion appears only in stars cooler 
than 6000 K. The candidate solar twins have Li abundance similar to that of our star, indicating that they also 
share with the Sun a similar mixing history. 

\begin{acknowledgements}
We are grateful to F. Castelli for helping us to understand how hydrogen profiles are computed in SYNTHE.
KB has been supported by the ESO DGDF 2006, and by the Italian {\em Ministero dell'Istruzione, Universit\`a e 
Ricerca} (MIUR) fundings. PB acknowledges support from EU contract MEXT-CT-2004-014265 
(CIFIST). This research has made use of SIMBAD and VIZIER databases, operated at CDS (Strasbourg, France).
\end{acknowledgements}

{}
\newpage

\appendix{

\section{On line material}

\begin{table*}  
\caption{Object ID, coordinates (Equinox J2000, Epoch J2000.13), photometry, and proper motions of the targets 
(see \citealt{Yadav2008} for details). In the last column the name of the stars according to 
\cite{sanders1977} or to \cite{Montgomery1993} are given. For a few stars two names are given, because both 
stars are within 1 arc sec circle, according to SIMBAD.}
\label{tab:total_targets}
\scriptsize
\begin{center}
\begin{tabular}{lcccccrrl}
\hline
\hline
Object  & $\alpha$	& $\delta$  &  $B\pm\Delta B$  &  $V\pm\Delta V$  &  $I\pm\Delta I$  & $\mu_x\pm\Delta \mu_x$ & $\mu_y\pm\Delta \mu_y$ & Name\\
 & (\degr) & (\degr) & (mag) & (mag) & (mag) & (mas/yr) & (mas/yr)\\ 
\hline
Obj219  & 132.893537 & 11.632623 & 13.715$\pm$0.009 & 13.100$\pm$0.008 & 12.425$\pm$0.005 &    1.96$\pm$1.31  &    8.63$\pm$3.03 & S1197\\
Obj266  & 132.860339 & 11.643584 & 14.212$\pm$0.006 & 13.601$\pm$0.006 & 12.939$\pm$0.007 & $-$0.24$\pm$2.38  &    0.18$\pm$3.45 & S944\\
Obj285  & 132.849450 & 11.647816 & 15.165$\pm$0.006 & 14.461$\pm$0.000 & 13.713$\pm$0.002 & $-$0.42$\pm$1.78  & $-$2.80$\pm$1.37 & S945\\
Obj288  & 132.900657 & 11.648980 & 14.494$\pm$0.013 & 13.857$\pm$0.004 & 13.160$\pm$0.005 &    0.00$\pm$0.77  &    1.61$\pm$2.44 & S1201\\
Obj291  & 132.937203 & 11.649712 & 14.090$\pm$0.010 & 13.478$\pm$0.009 & 12.807$\pm$0.018 &    0.18$\pm$1.13  &    7.26$\pm$3.03  & S1202\\
Obj342  & 132.823311 & 11.660013 & 14.118$\pm$0.004 & 13.503$\pm$0.005 & 12.841$\pm$0.008 &    0.77$\pm$2.14  &    5.59$\pm$4.22  & S948\\
Obj349  & 132.979809 & 11.661163 & 14.978$\pm$0.006 & 14.301$\pm$0.002 & 13.614$\pm$0.007 &    2.92$\pm$8.87  &    2.92$\pm$4.82 & MMJ6241/S1423\\
Obj350  & 132.836621 & 11.661145 & 14.226$\pm$0.005 & 13.624$\pm$0.010 & 12.955$\pm$0.006 & $-$1.96$\pm$1.07  &    0.18$\pm$1.49 & S950 \\
Obj364  & 132.794912 & 11.664138 & 15.288$\pm$0.005 & 14.584$\pm$0.010 & 13.809$\pm$0.006 & $-$1.37$\pm$1.55  &    0.24$\pm$1.43 & S951 \\
Obj401  & 132.829348 & 11.671034 & 14.268$\pm$0.006 & 13.661$\pm$0.007 & 13.009$\pm$0.003 & $-$3.81$\pm$2.56  &    1.31$\pm$1.96 & S954 \\
Obj437  & 132.827827 & 11.676879 & 14.600$\pm$0.006 & 13.998$\pm$0.008 & 13.331$\pm$0.002 &    2.98$\pm$0.71  & $-$2.02$\pm$3.69 & S956 \\
Obj455  & 132.971225 & 11.681574 & 13.930$\pm$0.007 & 13.301$\pm$0.004 & 12.608$\pm$0.018 &    5.30$\pm$5.65  &    5.77$\pm$2.80 & S1428 \\
Obj473  & 132.809740 & 11.685891 & 15.142$\pm$0.011 & 14.443$\pm$0.008 & 13.731$\pm$0.004 &    1.31$\pm$1.13  & $-$1.55$\pm$3.57 & S958\\
Obj571  & 132.951947 & 11.706357 & 14.686$\pm$0.012 & 14.021$\pm$0.014 & 13.309$\pm$0.004 & $-$0.89$\pm$1.31  & $-$0.30$\pm$3.21 & S1211\\
Obj574  & 133.024207 & 11.706858 & 14.309$\pm$0.016 & 13.653$\pm$0.007 & 12.932$\pm$0.005 & $-$1.73$\pm$1.01  & $-$0.24$\pm$1.07 & MMJ6362/S1431\\
Obj587  & 132.911415 & 11.710387 & 14.753$\pm$0.006 & 14.107$\pm$0.006 & 13.405$\pm$0.020 & $-$0.89$\pm$0.83  &    1.31$\pm$3.15 & S1213\\
Obj613  & 132.825417 & 11.715202 & 13.907$\pm$0.015 & 13.254$\pm$0.006 & 12.594$\pm$0.022 &    1.78$\pm$2.62  & $-$0.89$\pm$3.69 & S964\\
Obj637  & 132.840991 & 11.721590 & 15.191$\pm$0.017 & 14.489$\pm$0.010 & 13.751$\pm$0.011 &    0.65$\pm$0.54  & $-$0.42$\pm$2.50 & S966\\
Obj673  & 132.722266 & 11.727791 & 15.062$\pm$0.002 & 14.356$\pm$0.009 & 13.569$\pm$0.001 &    0.00$\pm$0.77  &    0.24$\pm$2.86 & S746 \\
Obj681  & 132.773982 & 11.729705 & 14.715$\pm$0.006 & 14.018$\pm$0.016 & 13.254$\pm$0.008 &    1.19$\pm$1.31  &    1.49$\pm$2.38 & S747\\
Obj689  & 132.875550 & 11.730521 & 13.783$\pm$0.012 & 13.120$\pm$0.011 & 12.436$\pm$0.010 & $-$3.21$\pm$2.86  & $-$0.36$\pm$3.87 & S1219\\
Obj713  & 132.840699 & 11.734734 & 14.886$\pm$0.017 & 14.172$\pm$0.009 & 13.421$\pm$0.017 &    1.55$\pm$1.37  & $-$0.48$\pm$1.31 & S969 \\
Obj750  & 132.722677 & 11.742964 & 14.215$\pm$0.007 & 13.576$\pm$0.014 & 12.865$\pm$0.005 &    0.18$\pm$2.08  &    1.07$\pm$2.20 & S750 \\
Obj756  & 132.947493 & 11.745099 & 15.393$\pm$0.011 & 14.695$\pm$0.008 & 13.923$\pm$0.009 &    3.03$\pm$0.89  &    5.06$\pm$1.73 & S1222 \\
Obj769  & 132.959247 & 11.749185 & 14.119$\pm$0.004 & 13.478$\pm$0.003 & 12.771$\pm$0.005 & $-$0.36$\pm$1.13  &    0.12$\pm$2.98 & S1224a\\
Obj778  & 132.836684 & 11.750684 & 13.716$\pm$0.010 & 13.093$\pm$0.011 & 12.411$\pm$0.011 &    1.37$\pm$1.78  &    2.98$\pm$4.40 & S976 \\
Obj809  & 132.758539 & 11.755306 & 15.696$\pm$0.007 & 14.959$\pm$0.015 & 14.162$\pm$0.004 & $-$1.13$\pm$1.61  & $-$1.49$\pm$1.84 & S754 \\
Obj851  & 132.854122 & 11.761931 & 14.730$\pm$0.007 & 14.113$\pm$0.004 & 13.449$\pm$0.007 &    0.71$\pm$3.09  &    2.98$\pm$1.07 & S982 \\
Obj880  & 132.770112 & 11.765793 & 14.202$\pm$0.009 & 13.544$\pm$0.014 & 12.841$\pm$0.002 & $-$0.30$\pm$1.07  & $-$1.07$\pm$1.61 & S757 \\
Obj905  & 132.746798 & 11.770272 & 14.038$\pm$0.004 & 13.434$\pm$0.010 & 12.748$\pm$0.015 &    1.01$\pm$1.73  & $-$1.73$\pm$2.74 & S758\\
Obj911  & 132.791312 & 11.771367 & 15.220$\pm$0.006 & 14.547$\pm$0.010 & 13.785$\pm$0.010 &    0.36$\pm$1.31  &    1.19$\pm$1.31 & S991\\
Obj917  & 133.021288 & 11.772611 & 15.497$\pm$0.007 & 14.755$\pm$0.008 & 13.923$\pm$0.006 & $-$0.54$\pm$2.56  &    2.44$\pm$2.62  & S1442\\
Obj971  & 132.884919 & 11.779352 & 15.287$\pm$0.005 & 14.592$\pm$0.004 & 13.793$\pm$0.000 & $-$0.12$\pm$0.59  &    0.36$\pm$0.30 & S1246\\
Obj986  & 132.905212 & 11.782141 & 14.646$\pm$0.008 & 14.007$\pm$0.003 & 13.283$\pm$0.001 &    0.24$\pm$1.25  &    1.19$\pm$0.54 & S1247 \\
Obj988  & 132.892171 & 11.782156 & 14.819$\pm$0.005 & 14.180$\pm$0.001 & 13.475$\pm$0.002 & $-$0.65$\pm$0.65  & $-$0.24$\pm$1.49 & S1248 \\
Obj1010 & 132.850486 & 11.785927 & 14.104$\pm$0.007 & 13.478$\pm$0.009 & 12.781$\pm$0.021 & $-$0.36$\pm$0.95  &    3.57$\pm$1.43 & S2209\\
Obj1032 & 132.985778 & 11.790265 & 14.997$\pm$0.005 & 14.358$\pm$0.003 & 13.649$\pm$0.001 &    3.33$\pm$3.33  & $-$0.24$\pm$0.48 & S1449 \\
Obj1036 & 132.851246 & 11.791211 & 15.678$\pm$0.018 & 14.947$\pm$0.003 & 14.164$\pm$0.005 & $-$0.36$\pm$0.24  & $-$1.01$\pm$1.25 & S1004 \\
Obj1051 & 132.922899 & 11.793371 & 14.726$\pm$0.007 & 14.090$\pm$0.004 & 13.382$\pm$0.004 & $-$1.55$\pm$0.54  &    1.96$\pm$1.07 & S1252 \\
Obj1062 & 132.900316 & 11.796392 & 15.144$\pm$0.001 & 14.477$\pm$0.008 & 13.745$\pm$0.004 &    1.25$\pm$1.07  & $-$0.12$\pm$2.68 & S1255 \\
Obj1067 & 133.014592 & 11.796693 & 15.201$\pm$0.005 & 14.559$\pm$0.009 & 13.824$\pm$0.002 &    0.24$\pm$2.08  & $-$0.30$\pm$1.25 & S1452 \\
Obj1075 & 132.912716 & 11.798715 & 14.386$\pm$0.000 & 13.712$\pm$0.006 & 12.992$\pm$0.004 &    1.19$\pm$1.96  & $-$2.80$\pm$3.33 & S1256 \\
Obj1088 & 132.953942 & 11.800602 & 15.151$\pm$0.004 & 14.492$\pm$0.001 & 13.760$\pm$0.007 &    0.77$\pm$1.90  & $-$3.99$\pm$2.56 & S1258 \\
Obj1090 & 132.869575 & 11.800607 & 14.450$\pm$0.004 & 13.800$\pm$0.005 & 13.040$\pm$0.010 &    0.18$\pm$1.73  & $-$1.55$\pm$1.07  & S1011\\
Obj1091 & 132.853518 & 11.801110 & 15.237$\pm$0.011 & 14.513$\pm$0.007 & 13.669$\pm$0.008 &    0.12$\pm$0.83  & $-$0.48$\pm$1.55  & S1012 \\
Obj1101 & 133.035214 & 11.802908 & 15.377$\pm$0.001 & 14.675$\pm$0.004 & 13.903$\pm$0.001 & $-$0.06$\pm$1.31  & $-$2.08$\pm$2.50  & MMJ6384 \\
Obj1108 & 132.855293 & 11.803762 & 14.878$\pm$0.005 & 14.177$\pm$0.001 & 13.386$\pm$0.006 & $-$0.30$\pm$0.95  & $-$0.42$\pm$0.71 & S1014 \\
Obj1129 & 132.885730 & 11.806694 & 14.795$\pm$0.005 & 14.171$\pm$0.006 & 13.482$\pm$0.002 &    1.13$\pm$0.30  & $-$1.55$\pm$1.31 & S1260 \\
Obj1137 & 132.800104 & 11.807413 & 15.571$\pm$0.002 & 14.873$\pm$0.008 & 14.107$\pm$0.009 & $-$1.90$\pm$0.89  &    0.83$\pm$1.19 & S2213 \\
Obj1161 & 132.981615 & 11.810602 & 14.549$\pm$0.004 & 13.883$\pm$0.010 & 13.149$\pm$0.006 &    0.89$\pm$4.46  & $-$6.84$\pm$2.86 & S1457 \\
Obj1163 & 132.787098 & 11.810483 & 14.688$\pm$0.009 & 13.992$\pm$0.004 & 13.221$\pm$0.002 &    0.42$\pm$0.65  &    8.03$\pm$1.19 & S1022 \\
Obj1194 & 132.753356 & 11.814655 & 15.281$\pm$0.002 & 14.614$\pm$0.010 & 13.876$\pm$0.002 &    0.30$\pm$1.01  & $-$0.42$\pm$0.65 & S770 \\
Obj1197 & 132.877372 & 11.815215 & 13.921$\pm$0.006 & 13.315$\pm$0.003 & 12.618$\pm$0.024 & $-$0.18$\pm$1.07  &    1.78$\pm$0.77 & MMJ5882/S1264b\\
Obj1247 & 132.812708 & 11.822522 & 14.753$\pm$0.008 & 14.144$\pm$0.008 & 13.470$\pm$0.005 &    0.65$\pm$1.25  & $-$2.80$\pm$1.96 & S1033 \\
Obj1303 & 132.736097 & 11.831844 & 15.318$\pm$0.008 & 14.641$\pm$0.008 & 13.899$\pm$0.009 &    0.06$\pm$2.26  &    0.89$\pm$1.25 & S779 \\
Obj1304 & 132.858515 & 11.832075 & 15.454$\pm$0.015 & 14.731$\pm$0.009 & 13.916$\pm$0.006 & $-$0.12$\pm$1.31  & $-$0.12$\pm$1.90 & S1041\\
Obj1315 & 132.994897 & 11.834025 & 14.990$\pm$0.013 & 14.297$\pm$0.011 & 13.544$\pm$0.008 & $-$1.19$\pm$4.40  &    0.48$\pm$2.20 & MMJ6306/S1462 \\
Obj1334 & 132.866453 & 11.836636 & 15.083$\pm$0.007 & 14.403$\pm$0.007 & 13.669$\pm$0.000 &    0.77$\pm$1.61  & $-$1.84$\pm$0.89 & S1048\\
Obj1342 & 132.826093 & 11.838787 & 14.935$\pm$0.006 & 14.285$\pm$0.005 & 13.547$\pm$0.001 &    1.37$\pm$1.49  & $-$0.36$\pm$0.06 & S1050 \\
Obj1387 & 132.874854 & 11.852505 & 14.724$\pm$0.004 & 14.098$\pm$0.002 & 13.398$\pm$0.000 &    0.59$\pm$3.75  & $-$1.55$\pm$7.97 & S1283 \\
Obj1392 & 132.749200 & 11.853500 & 15.527$\pm$0.002 & 14.811$\pm$0.004 & 14.047$\pm$0.000 & $-$1.49$\pm$2.08  & $-$1.61$\pm$0.77 & S785 \\
Obj1397 & 132.882992 & 11.854616 & 14.631$\pm$0.001 & 14.009$\pm$0.003 & 13.304$\pm$0.001 &    1.84$\pm$2.32  & $-$3.03$\pm$5.89 & S1287 \\
Obj1424 & 132.890226 & 11.862493 & 13.825$\pm$0.009 & 13.203$\pm$0.005 & 12.501$\pm$0.000 &    1.49$\pm$2.50  & $-$1.25$\pm$3.87 & Check  \\
Obj1458 & 132.762475 & 11.873808 & 15.716$\pm$0.010 & 14.977$\pm$0.005 & 14.186$\pm$0.004 &    1.01$\pm$1.84  &    0.12$\pm$1.49 & S795\\
Obj1480 & 132.916988 & 11.878716 & 14.383$\pm$0.008 & 13.783$\pm$0.008 & 13.125$\pm$0.002 & $-$3.99$\pm$2.50  &    1.25$\pm$2.32 & S1300\\
Obj1496 & 132.781325 & 11.882262 & 14.486$\pm$0.013 & 13.879$\pm$0.004 & 13.214$\pm$0.002 & $-$0.95$\pm$0.12  & $-$0.30$\pm$0.65 & S2224 \\
Obj1504 & 132.864557 & 11.884028 & 14.796$\pm$0.001 & 14.171$\pm$0.011 & 13.474$\pm$0.000 & $-$1.13$\pm$0.95  &    1.31$\pm$1.31 & S1078 \\
Obj1514 & 132.753181 & 11.886527 & 15.498$\pm$0.003 & 14.777$\pm$0.004 & 14.008$\pm$0.000 & $-$0.12$\pm$1.13  &    1.73$\pm$1.37 & S802 \\
Obj1587 & 132.846438 & 11.901394 & 14.804$\pm$0.015 & 14.163$\pm$0.004 & 13.469$\pm$0.006 & $-$0.95$\pm$1.73  & $-$1.01$\pm$1.78 & S1087 \\
Obj1622 & 132.801216 & 11.906389 & 14.788$\pm$0.004 & 14.156$\pm$0.002 & 13.459$\pm$0.004 &    1.43$\pm$1.25  &    0.95$\pm$0.65 & S1089 \\
Obj1680 & 132.883591 & 11.919069 & 14.291$\pm$0.015 & 13.646$\pm$0.006 & 12.951$\pm$0.004 &    1.01$\pm$1.13  &    0.18$\pm$1.07& S1314 \\
Obj1706 & 133.010317 & 11.926166 & 15.468$\pm$0.005 & 14.745$\pm$0.017 & 13.981$\pm$0.006 &    0.54$\pm$0.95  &    0.30$\pm$0.54 & MMJ6341/S1481\\
Obj1716 & 132.866205 & 11.928044 & 13.918$\pm$0.010 & 13.299$\pm$0.009 & 12.625$\pm$0.005 & $-$1.96$\pm$1.84  &    4.82$\pm$3.15 & S1092\\
Obj1722 & 132.828014 & 11.930478 & 14.731$\pm$0.002 & 14.130$\pm$0.006 & 13.449$\pm$0.009 & $-$3.57$\pm$26.24 & $-$2.44$\pm$2.20 & S1093 \\
Obj1735 & 133.001728 & 11.935288 & 14.993$\pm$0.012 & 14.332$\pm$0.010 & 13.617$\pm$0.007 &    1.96$\pm$1.78  & $-$2.92$\pm$1.55 & S1483 \\
Obj1758 & 132.746824 & 11.943570 & 13.860$\pm$0.056 & 13.207$\pm$0.015 & 12.545$\pm$0.007 & $-$0.89$\pm$0.48  &    3.45$\pm$1.67 & MMJ5342/S816 \\
Obj1768 & 132.906577 & 11.945718 & 15.060$\pm$0.003 & 14.404$\pm$0.004 & 13.684$\pm$0.004 &    0.77$\pm$1.13  & $-$1.01$\pm$0.89 & MMJ6028/S1320\\
Obj1778 & 132.737778 & 11.947424 & 15.679$\pm$0.004 & 14.948$\pm$0.004 & 14.155$\pm$0.003 &    0.83$\pm$1.96  & $-$0.65$\pm$0.65 & MMJ5310/S820\\
Obj1787 & 132.788102 & 11.950097 & 15.214$\pm$0.006 & 14.547$\pm$0.004 & 13.813$\pm$0.003 &    1.61$\pm$1.73  &    2.26$\pm$2.26  & MMJ5484\\
Obj1788 & 132.804106 & 11.950254 & 15.104$\pm$0.008 & 14.441$\pm$0.004 & 13.709$\pm$0.000 & $-$1.78$\pm$1.13  &    4.64$\pm$1.84 & MMJ5541\\
Obj1842 & 132.786897 & 11.964913 & 14.844$\pm$0.007 & 14.237$\pm$0.002 & 13.557$\pm$0.005 & $-$0.36$\pm$1.01  & $-$1.61$\pm$1.49 & MMJ5479/S1102\\
Obj1852 & 133.013763 & 11.967948 & 14.575$\pm$0.011 & 13.962$\pm$0.008 & 13.286$\pm$0.005 & $-$1.01$\pm$2.32  &    1.67$\pm$3.51 & S1486 \\
Obj1862 & 132.743111 & 11.970785 & 15.126$\pm$0.001 & 14.483$\pm$0.003 & 13.753$\pm$0.004 & $-$0.06$\pm$1.37  &    0.89$\pm$1.13 & MMJ5331\\
Obj1903 & 132.783150 & 11.981489 & 15.422$\pm$0.004 & 14.733$\pm$0.003 & 13.971$\pm$0.001 & $-$1.67$\pm$1.25  &    0.83$\pm$0.83 & MMJ5469\\
Obj1948 & 132.928314 & 11.991953 & 14.627$\pm$0.009 & 14.015$\pm$0.004 & 13.327$\pm$0.002 &   0.42 $\pm$1.67  & $-$3.45$\pm$2.98 & S1330 \\
Obj1955 & 132.743729 & 11.994304 & 14.842$\pm$0.001 & 14.212$\pm$0.004 & 13.483$\pm$0.002 &   0.59 $\pm$0.54  & $-$2.98$\pm$3.39  & MMJ5338/S829 \\
Obj1957 & 132.888539 & 11.994779 & 14.406$\pm$0.007 & 13.789$\pm$0.004 & 13.085$\pm$0.008 & $-$1.19$\pm$1.90  &    0.18$\pm$3.69 & MMJ5962/S1331\\
Obj2016 & 132.931975 & 12.015297 & 14.158$\pm$0.002 & 13.553$\pm$0.007 & 12.841$\pm$0.000 &   0.77 $\pm$2.38  &    0.30$\pm$2.44 & S1333 \\
Obj2017 & 132.943039 & 12.015413 & 15.509$\pm$0.003 & 14.857$\pm$0.001 & 14.109$\pm$0.001 & $-$2.68$\pm$2.62  &    6.90$\pm$3.87 & S1334 \\
Obj2018 & 132.914220 & 12.015883 & 15.237$\pm$0.000 & 14.565$\pm$0.005 & 13.832$\pm$0.007 & $-$2.50$\pm$1.55  &    1.67$\pm$2.02 & MMJ6055 \\
\hline
\end{tabular}
\end{center}
\end{table*}
\normalsize

\begin{table*}  
\caption{Radial velocities, effective temperatures, and lithium abundances of the 59 stars retained as possible single members. The values in Table 
are derived by adding the difference of $T_{\rm eff}$ between the stars and the Sun to the solar temperature (5777 K). The solar temperature derived 
from the GIRAFFE spectrum and our calibrations is of 5792 and 5717 K for the LDR and H$\alpha$ methods, respectively. The 10 best solar twins 
candidates are indicated in bold face. The $S/N$ ratio/pixel of the co-added spectra varies between 80 and 110, depending on the magnitude of the stars.}
\label{tab:targets_selected}
\scriptsize
\begin{center}
\begin{tabular}{lcclrrcr}
\hline
Object & $V_{\rm rad}\pm\Delta V_{\rm rad}$ & $T_{\rm eff}^{\rm LDR}\pm\Delta T_{\rm eff}^{\rm LDR}$ & $T_{\rm eff}^{\rm H\alpha}$ & $EW_{\rm Li+Fe}$ &
$\log EW_{\rm Li}$ & $\log N{\rm (Li)}^{\rm LTE}$ & $\log N{\rm (Li)}^{\rm NLTE}$ \\
       & (km s$^{-1}$) & (K)      & (K)                               & (m\AA) & (m\AA) &    &\\ 
\hline
Sun    &                 & 5777$\pm$27 & 5777 &  12.1  &     0.3&  0.8  &  0.8	   \\
219    &  33.85$\pm$0.28 & 6243$\pm$54 & 6110 &  22.7  &     1.2&  2.1  &  2.1     \\
266    &  32.95$\pm$0.46 & 6147$\pm$63 & 6060 &  53.2  &     1.7&  2.6  &  2.5     \\
{\bf 285 }   &  33.72$\pm$0.67 & 5836$\pm$67 & 5777 &   1.5  & $<$ 0.0&  0.6  &  $<$ 0.6 \\
288    &  34.26$\pm$0.40 & 6004$\pm$67 & 6010 &  42.4  &     1.5&  2.3  &  2.3     \\
291    &  32.13$\pm$0.29 & 6177$\pm$57 & 6160 &  59.6  &     1.7&  2.6  &  2.6     \\
349    &  34.28$\pm$0.42 & 5952$\pm$78 & 5917 &  10.4  & $<$ 0.0&  0.7  &  $<$ 0.7 \\
350    &  32.56$\pm$0.31 & 6024$\pm$52 & 6010 &  68.2  &     1.8&  2.6  &  2.6     \\
401    &  32.64$\pm$0.37 & 6165$\pm$64 & 6110 &  44.9  &     1.6&  2.5  &  2.4     \\
473    &  34.74$\pm$0.38 & 5919$\pm$76 & 5807 &   2.7  & $<$ 0.0&  0.7  &  $<$ 0.7 \\
587    &  33.06$\pm$0.39 & 6077$\pm$65 & 6060 &  37.2  &     1.4&  2.3  &  2.3     \\
613    &  33.05$\pm$0.29 & 6202$\pm$45 & 6110 &  61.3  &     1.7&  2.7  &  2.6     \\
{\bf 637  }  &  34.00$\pm$0.50 & 5806$\pm$65 & 5777 &  17.8  &     0.9&  1.4  &  1.4    \\
673    &  32.44$\pm$0.57 & 5639$\pm$63 & 5747 &  19.1  &     0.9&  1.4  &  1.4     \\
689    &  32.88$\pm$0.24 & 6093$\pm$41 & 6110 &  42.3  &     1.5&  2.4  &  2.3     \\
750    &  33.13$\pm$0.28 & 5918$\pm$48 & 5927 &  16.4  &     0.9&  1.5  &  1.5     \\
769    &  34.47$\pm$0.29 & 5984$\pm$46 & 6010 &  29.4  &     1.3&  2.1  &  2.0     \\
778    &  33.45$\pm$0.24 & 6114$\pm$39 & 6060 &  15.6  &     0.8&  1.7  &  1.7     \\
809    &  31.87$\pm$0.48 & 5667$\pm$78 & 5537 &   6.9  & $<$ 0.0&  0.4  &  $<$ 0.5 \\
851    &  33.47$\pm$0.31 & 5948$\pm$61 & 6060 &  12.9  &     0.6&  1.3  &  1.3     \\
911    &  32.08$\pm$0.35 & 5885$\pm$67 & 5837 &  13.4  &     0.6&  1.2  &  1.2     \\
988    &  32.07$\pm$0.28 & 5935$\pm$53 & 6060 &  28.2  &     1.3&  2.0  &  2.0     \\
1032   &  34.02$\pm$0.47 & 5955$\pm$60 & 6010 &  17.8  &     0.9&  1.6  &  1.6     \\
1036   &  33.39$\pm$0.48 & 5612$\pm$66 & 5537 &  20.9  &     1.0&  1.4  &  1.4     \\
1051   &  32.21$\pm$0.29 & 6081$\pm$55 & 6060 &  34.7  &     1.4&  2.2  &  2.2     \\
1062   &  32.64$\pm$0.45 & 5926$\pm$55 & 5867 &  21.0  &     1.1&  1.7  &  1.7     \\
1067   &  33.37$\pm$0.35 & 5929$\pm$69 & 5917 &  11.4  &     0.4&  1.0  &  1.0     \\
1075   &  33.13$\pm$0.24 & 5871$\pm$48 & 5917 &  10.7  &     0.0&  0.6	&  0.7	   \\
1088   &  32.87$\pm$0.29 & 5890$\pm$59 & 5867 &   9.3  & $<$ 0.0&  0.6  &  $<$ 0.7 \\
1090   &  33.34$\pm$0.32 & 6086$\pm$54 & 6010 &  41.2  &     1.5&  2.3  &  2.3	   \\
{\bf 1101}   &  32.72$\pm$0.34 & 5756$\pm$60 & 5717 &   6.9  & $<$ 0.0&  0.5  &  $<$ 0.6 \\
1129   &  33.92$\pm$0.30 & 5959$\pm$51 & 6010 &  32.4  &     1.4&  2.1  &  2.1     \\
1137   &  33.59$\pm$0.48 & 5741$\pm$69 & 5627 &   5.3  & $<$ 0.0&  0.5  &  $<$ 0.6 \\
{\bf 1194}   &  33.30$\pm$0.40 & 5766$\pm$64 & 5837 &   5.3  & $<$ 0.0&  0.5  &  $<$ 0.6 \\
1197   &  34.26$\pm$0.28 & 6207$\pm$44 & 6110 &  28.1  &     1.3&  2.2  &  2.2     \\
1247   &  32.44$\pm$0.51 & 5994$\pm$60 & 6010 &  30.2  &     1.3&  2.1  &  2.1     \\
{\bf 1303 }  &  32.65$\pm$0.41 & 5716$\pm$64 & 5717 &  10.1  & $<$ 0.0&  0.5  &  $<$ 0.6 \\
{\bf 1304  } &  33.60$\pm$0.39 & 5704$\pm$64 & 5717 &   7.9  & $<$ 0.0&  0.5  &  $<$ 0.5 \\
{\bf 1315   }&  32.55$\pm$0.34 & 5874$\pm$58 & 5867 &  15.6  &     0.8&  1.4  &  1.4    \\
1334   &  32.37$\pm$0.45 & 5957$\pm$57 & 5957 &  29.2  &     1.3&  2.0  &  2.0     \\
1387   &  33.35$\pm$0.24 & 6090$\pm$58 & 6060 &  37.5  &     1.5&  2.3  &  2.3     \\
{\bf 1392}   &  33.68$\pm$0.57 & 5716$\pm$63 & 5687 &   6.3  & $<$ 0.0&  0.5  &  $<$ 0.6 \\
1458   &  32.55$\pm$0.56 & 5640$\pm$65 & 5567 &   6.1  & $<$ 0.0&  0.4  &  $<$ 0.5 \\
1496   &  34.22$\pm$0.48 & 6173$\pm$54 & 6160 &  63.1  &     1.7&  2.7  &  2.6     \\
1504   &  33.39$\pm$0.48 & 5934$\pm$56 & 6060 &  40.7  &     1.5&  2.2  &  2.2     \\
1514   &  33.33$\pm$0.49 & 5613$\pm$67 & 5597 &  25.6  &     1.2&  1.6  &  1.6     \\
1587   &  32.38$\pm$0.62 & 5975$\pm$55 & 6010 &  28.6  &     1.3&  2.0  &  2.0     \\
1622   &  33.26$\pm$0.61 & 6043$\pm$57 & 6010 &  34.6  &     1.4&  2.2  &  2.2     \\
1716   &  33.84$\pm$0.59 & 6030$\pm$40 & 6060 &  40.8  &     1.5&  2.3  &  2.3     \\
1722   &  33.87$\pm$0.52 & 6007$\pm$62 & 6010 &  38.1  &     1.5&  2.3  &  2.2     \\
1735   &  33.18$\pm$0.59 & 5959$\pm$59 & 5960 &  10.9  &     0.2&  0.9  &  0.9     \\
1758   &  33.78$\pm$0.67 & 6221$\pm$52 & 6160 &  36.3  &     1.4&  2.4  &  2.4     \\
1768   &  33.98$\pm$0.57 & 5844$\pm$57 & 5927 &  40.1  &     1.5&  2.1  &  2.1     \\
{\bf 1787 }  &  33.44$\pm$0.57 & 5768$\pm$70 & 5807 &  12.9  &     0.5&  1.1  &  1.0   \\
1788   &  33.38$\pm$0.80 & 5886$\pm$60 & 5867 &  23.4  &     1.1&  1.8  &  1.8     \\
1852   &  32.30$\pm$0.40 & 6009$\pm$63 & 6010 &  32.2  &     1.4&  2.1  &  2.1     \\
1903   &  32.76$\pm$0.42 & 5609$\pm$72 & 5687 &   6.8  & $<$ 0.0&  0.4  &  $<$ 0.5 \\
1948   &  32.86$\pm$0.29 & 6164$\pm$63 & 6010 &  43.0  &     1.5&  2.4  &  2.4     \\
1955   &  32.63$\pm$0.60 & 5961$\pm$76 & 5837 &  35.8  &     1.4&  2.2  &  2.2     \\
{\bf 2018}   &  31.78$\pm$0.43 & 5693$\pm$74 & 5777 &   8.7  & $<$ 0.0&  0.4  &  $<$ 0.5 \\
\hline
\end{tabular}
\end{center}
\end{table*}

\begin{table*}  
\caption{Radial velocities of likely binaries or non members. For these stars each of the three RV measurements 
is given. Fields with blanks indicate that problems were present with the cross-correlation profile of these 
objects.}
\label{tab:nonmembers}
\small
\begin{center}
\begin{tabular}{lrrcccc}
\hline
\hline
Object & $V_{\rm rad}^1$ & $\Delta V_{\rm rad}^1$ &  $V_{\rm rad}^2$ & $\Delta V_{\rm rad}^2$ & $V_{\rm rad}^3$ & $\Delta V_{\rm rad}^3$ \\ \hline
342	&   17.74  & 0.72 &   57.20	& 1.45 & 40.75  &  0.73	     \\
364	&  36.93   & 0.43 &   36.58     & 1.18 & 36.48	&  0.74	    \\
437	&  45.87   & 0.65 &   45.54     & 0.83 & 45.68  &  0.52	    \\
455	&  38.07   & 0.50 &   14.62     & 1.24 & 33.41	&  0.47	    \\
571	&  31.73   & 0.49 &   31.07     & 0.66 & 31.47  &  0.56	    \\
574	&  42.88   & 0.46 &   42.91     & 0.85 & 42.08  &  0.62	    \\
681	&  24.14   & 0.80 &   56.70	& 1.22 & 59.70  &  0.85	     \\
713	&  35.34   & 0.46 &   35.03     & 0.64 & 34.91  &  0.33	     \\
756	&  46.24   & 0.50 &   45.54     & 0.77 & 45.47  &  0.41	     \\
880	&  39.62   & 6.67 &     /	&   /  & 33.55  &  5.81	   \\
905	&  34.78   & 0.35 &   37.30	& 1.02 & 74.65  &  0.89    \\
917	&  18.66   & 0.86 &   15.59	& 1.36 &  /     &   /	       \\
971	&  36.13   & 0.51 &   36.13	& 0.70 & 36.22  &  0.63     \\
986	&  16.54   & 0.53 &   27.69	& 0.84 & 49.08  &  0.68     \\
1010	&  17.22   & 0.97 &   30.77	& 0.63 & 34.29  &  0.75	 \\
1091	&  27.59   & 3.27 &   27.03	& 0.62 & 25.77  &  0.78	 \\
1108	&  15.92   & 1.00 &     /       & /    &  /     & /	 \\
1161	&  35.53   & 0.65 &   35.49	& 1.21 & 22.45	&  1.55 	 \\
1163	&  10.10   & 0.66 &   65.99	& 1.32 & 70.80	&  1.24 	 \\
1342	&  37.97   & 0.69 &   28.40	& 1.06 & 34.93  &  0.65  	 \\
1397	&  35.43   & 1.12 &   34.91	& 0.85 & 34.88  &  0.54 	  \\
1424	&  54.44   & 1.62 &   57.22	& 1.31 & 56.04  &  1.58	    \\
1480	&  22.86   & 1.02 &   25.56	& 1.09 & 28.27  &  0.48	    \\
1680	&  37.65   & 1.29 &   35.65	& 0.96 & 32.27  &  0.81      \\
1706	&  35.79   & 1.05 &   35.40	& 0.90 & 35.77  &  0.61      \\
1778	&  29.59   & 1.30 &   29.92	& 2.96 & 31.19  &  1.26 	  \\
1842	&  30.09   & 0.69 &   29.88	& 1.20 & 30.52  &  0.47    \\
1862	&  31.31   & 0.63 &   30.29	& 2.03 & 31.27  &  0.43        \\
1957	&  21.17   & 0.52 &   20.15	& 0.83 & 21.21  &  0.57 \\
2016	&  21.64   & 0.46 &   25.96	& 0.94 & 65.63  &  0.66 \\
2017	& 126.58   & 0.84 &   125.09~~	& 1.62 &126.01~~  &  0.63  	\\
\hline
\end{tabular}
\end{center}
\end{table*}

}

\end{document}